# Elementary excitations and crossover phenomenon in liquids

T. Iwashita[1], D. M. Nicholson[2] and T. Egami[1,2,3]

[1]*Joint Institute for Neutron Sciences and Department of Physics and Astronomy, University of Tennessee, Knoxville, TN 37996*

[2]*Oak Ridge National Laboratory, Oak Ridge, TN 37831*

[3]*Department of Materials Science and Engineering, University of Tennessee, Knoxville, TN 37996 USA*

**Abstract**

The elementary excitations of vibration in solids are phonons. But in liquids phonons are extremely short-lived and marginalized. In this letter through classical and *ab-initio* molecular dynamics simulations of the liquid state of various metallic systems we show that different excitations, the local configurational excitations in the atomic connectivity network, are the elementary excitations in high temperature metallic liquids. We also demonstrate that the competition between the configurational excitations and phonons determines the so-called crossover phenomenon in liquids. These discoveries open the way to the explanation of various complex phenomena in liquids, such as fragility and the rapid increase in viscosity toward the glass transition, in terms of these excitations.

PACS#:

The physics of phonon in crystalline solids is well-understood [1]. For instance its frequency can be calculated by diagonalizing the dynamical (Hessian) matrix, the matrix of which elements are the second derivative of the total energy with respect to the atomic displacements [1]. In liquids, however, the dynamical matrix itself is time-dependent because the atomic structure is changing with time. If the dynamical matrix varies as fast as, or faster than phonons [2] the validity of the concept of phonon becomes questionable. Yet, it is known that at high temperatures the specific heat of a liquid approximately satisfies the Dulong-Petit law of $C_V = 3k_B$ [3], suggesting the presence of some well-defined excitations. But the nature of such excitations in liquids is unknown. In general we know very little about the atomic dynamics in liquids. Theoretically the flow of a liquid is commonly described by continuum hydrodynamic theories [4, 5]. Less attention has been paid to the atomic level dynamics because it has been believed that the liquids were so random that details of the atomic motion were irrelevant to the physics of liquid flow. For instance the energy landscape theory describes atoms in a liquid at high temperatures as diffusing almost freely well above the peaks and valleys of energy landscape, seeing only shallow minima [6]. Actually at high temperatures above the so-called crossover temperature viscosity shows the Arrhenius behavior of thermal activation [7, 8]. However the nature of such activation processes and the origin of the crossover behavior remain unclear.

In this letter we show that elementary excitations different from phonons exist in liquids, and they are the origin of viscosity, and that the competition between the high temperature elementary excitations and phonons results in the crossover phenomenon in liquids. This is done through classical as well as *ab-initio* molecular dynamics (MD) simulations on various metallic liquids in either NVT or NPT ensembles. The periodic boundary condition on a cubic box was imposed for all systems. For liquid iron the model consisted of 16,384 atoms interacting via the modified Johnson potential [9]. The binary Kob-Andersen (KA) model [10] of 80/20 mixture (80% particle *A* and 20% particle *B*) consisted of 10,976 Lennard-Jones (LJ) particles at a reduced density of 1.2. All results for KA model are expressed in argon units. The simulations were performed in a temperature range from 60K to 300K, under NVT ensemble. For $Cu_{56}Zr_{44}$ and $Zr_{50}Cu_{40}Al_{10}$, we employed embedded atom method (EAM) potentials [11, 12] and simulated the systems with 16,000 atoms for $Cu_{56}Zr_{44}$ under NVT ensemble, and with 32,000 atoms for $Zr_{50}Cu_{40}Al_{10}$ under NPT ensemble with pressure equal to zero. The simulations were carried out using the LAMMPS software. *Ab initio* MD simulations of $Cu_{56}Zr_{44}$ in the NVT ensemble were performed using dynamics based on forces from the Projector Augmented-Wave (PAW) method [13] (Gamma point only) as implemented in the VASP 4.6 code [14]. The simulation cell contained 200 atoms with periodic boundary conditions. The sample was equilibrated using a Nosé-Hoover thermostat [15, 16] at $T$ = 4000, 4500 and 5000K for 3 ps followed by 1.1 ps of data collection during which the positions and stress matrix were recorded every time step (fs). Further details of the methods are given in Supplementary Material.

The viscosity, $\eta$, is a key parameter describing the dynamics of liquids. It is given by the Green-Kubo formula [17],

$$\eta = \frac{V}{k_B T} \int_0^\infty \langle \sigma^{xy}(0) \sigma^{xy}(t) \rangle dt \qquad (1)$$

where $\sigma^{xy}$ is the *x-y* component of the stress tensor, *T* is the temperature and *V* is the volume of the liquid. The time-scale of liquid dynamics is determined by the Maxwell relaxation time $\tau_M$,

$$\tau_M = \frac{\eta}{G_\infty}, \qquad (2)$$

where

$$G_\infty = \frac{V}{k_B T} \langle (\sigma^{xy}(0))^2 \rangle \qquad (3)$$

is the high-frequency shear modulus [17]. If the time-scale of the experiment is shorter than $\tau_M$ the system behaves like a solid, whereas if it longer it behaves like a liquid. It has been suggested that $\tau_M$ is related to the α-relaxation time, $\tau_\alpha$ [18], and the bond-lifetime, $\tau_B$ [19]. The α-relaxation time is determined from the intermediate scattering function at the first peak of the structure function, $S(Q)$ [17]. In the metallic systems under consideration here chemical bonds are not well-defined. However, it is possible to define the atomic connectivity network by defining the "bond" between the nearest neighbor atoms. This is because the nearest and the second nearest neighbors are well separated in the atomic pair-density function (PDF). We define the distinction between the 1st and 2nd nearest neighbors by the minimum in the PDF between the 1st and 2nd peaks.

In Figure 1 we compare the temperature dependences of $\tau_M$, $\tau_\alpha$, and $\tau_B$ for liquid Fe. Indeed three relaxation times show similar temperature dependences, but they are different in magnitude. $\tau_\alpha$ is longer than $\tau_M$ by a factor of about three, and $\tau_B$ is even longer, by an order of magnitude, than $\tau_\alpha$ [19]. On the other hand the life-time of the state of local atomic connectivity, $\tau_{LC}$, [20] is in much closer agreement with $\tau_M$ as shown also in Figure 1. Here $\tau_{LC}$ is defined as the time for an atom to lose or gain one nearest neighbor (Figure 2), thus changing the local atomic connectivity through local configurational excitation (LCE). In Figure 3 $\tau_M$ and $\tau_{LC}$ are shown in logarithmic scale against $1/T$ for various systems. Both show the Arrhenian behavior at high temperatures, but deviate from it below $T_A$, the crossover temperature. Figure 3 suggests that $\tau_M$ and $\tau_{LC}$ are similar in value above $T_A$ for all systems.

It is interesting to note that in all liquids, other than the KA model which has a rather different energy scale, $\tau_M$ and $\tau_{LC}$ extrapolate to about $\tau_\infty = 20$ fs at the limit of $T \rightarrow \infty$ (Table 1). This value of $\tau_\infty$ corresponds to about 200 meV. The activation energy is also about 200 meV. These results are more consistent with the LCE's characterized by the bond energy (250 meV for Fe) than phonons (the Debye frequency is about 40 meV).

We show in Figure 4 the ratio, $\tau_M/\tau_{LC}$, as a function of $T/T_A$ for various systems including the results of the *ab-initio* MD simulations. In spite of differences in composition and the method of MD this ratio, $\tau_M/\tau_{LC}$, is remarkably close to unity at temperatures above $T_A$. Given some small uncertainty in defining the nearest neighbors, as discussed in the Supplementary Material, the result strongly suggests that the relation,

$$\tau_M = \tau_{LC} \qquad (T > T_A), \qquad (4)$$

is universally valid for metallic liquids at high temperatures. This is an important result, because this equation directly connects a macroscopic quantity, $\tau_M$, with a microscopic quantity, $\tau_{LC}$. In liquids phonons are strongly scattered, short-lived, and cannot be used as the basis to explain other properties such as viscosity which shows the Arrhenian behavior. Instead, this result suggests that LCE is the elementary excitation in the high-temperature liquid. LCE's are not long-living harmonic excitations as are phonons. However, they are the elementary steps to change the atomic connectivity network, and directly control the macroscopic viscosity. Thus it is likely that LCE's represent the full degrees of freedom in the liquid as elementary excitations, just as phonons do in solids.

From equations (1) - (3) we obtain [21],

$$\tau_M = \int_0^\infty \frac{\langle \sigma^{xy}(0)\sigma^{xy}(t)\rangle}{\langle (\sigma^{xy}(0))^2\rangle} dt. \qquad (5)$$

Therefore equation (4) means that the lifetime of the shear correlation function is determined by the lifetime of local atomic connectivity. This is reasonable because the macroscopic shear stress is a sum of the atomic-level shear stresses [22], and the atomic-level stress is determined largely by the topology of the nearest neighbors [23]. For instance the atomic-level pressure is linearly related to the local coordination number [24]. Thus LCE's are the elementary excitations for the atomic-level stresses. This explains why the equipartition theorem for atomic-level stresses [23] as well as the Dulong-Petit law [3] are valid for high-temperature liquids.

We find that the crossover phenomena at $T_A$ are closely related to localization of phonons. The mean-free path of transverse phonon,

$$\xi_p = c_T \tau_M \qquad (6)$$

where $c_T$ ($= \sqrt{G_\infty/\rho_0}$, $\rho_0$ is the physical density) is the transverse sound velocity, decreases with increasing temperature. At high temperatures it is shorter than $a$, the distance between the nearest neighbors, and phonons are localized [25,26]. The temperature $T_I$, at which $\xi_p$ becomes equal to $a$ (see Supplementary Material), is shown by arrows in Figure 4, demonstrating that $T_I$ is very close to $T_A$. Agreement between $T_A$ and $T_I$ is not perfect, but this may be partly because $T_A$ is not so accurately defined due to the crossover nature of the phenomena. Interestingly in the case of KA glass we had to use the second neighbor B-B distance for $a$, because of the strong A-B chemical bond. Above $T_A$ the lifetime of local atomic configuration is so short that the local atomic connectivity changes before atoms communicate with neighbors through atomic vibrations. Thus LCE's are independent of each other above $T_A$.

In the energy landscape picture, below the crossover temperature the system starts to sample deeper landscape minima, and enters the "landscape influenced" regime [3, 27]. A number of properties of liquid exhibit significant changes in their nature through $T_A$ [6-8, 28, 29]. However, it had not been known what controlled the crossover point. The present result makes clear that the dynamic communication among atoms is the deciding factor, and the Ioffe-Regel localization [25] of transverse phonons into the LCE's is the physics behind this phenomenon.

Below $T_A$ the ratio, $\tau_M/\tau_{LC}$, increases with decreasing temperature. Because the local atomic connectivity is closely related to the atomic level stresses, an LCE, the action of breaking or creating a bond, will change the atomic level stresses of the two atoms involved in the broken or created bond. At $T > T_A$ $\tau_{LC}$ is too short for LCE to affect neighboring atoms other than the two involved in the bond, but at low temperatures $\tau_{LC}$ is long enough for LCE's to create dynamic long-range stress field around them, just as the atomic level stresses do [23, 24]. Thus LCE's can interact through the dynamic elastic field they create [30]. For instance they could shield each other to cancel the long-range elastic field to reduce the elastic energy. The increase in the $\tau_M/\tau_{LC}$ ratio below $T_A$ may be related to such interactions among LCE's. The definition of LCE as the elementary excitation in the high-temperature liquid state opens the possibility of describing the dynamics of low-temperature liquids in terms of interactions among LCE's. The way the ratio, $\tau_M/\tau_{LC}$, increases below $T_A$ is related to fragility [7]. Therefore fragility as well as the approach to the glass transition could be explained in terms of the interactions among LCE's. Incidentally, when the liquid flow is induced by applied stress below $T_g$, bond creation and annihilation are locally coupled to form a bond-exchange action, so that the $\tau_M/\tau_{LC}$ ratio is equal to 1/2 [20].

In conclusion we studied the dynamics of simple metallic liquids through classical as well as *ab-initio* MD simulations. We discovered that the dynamics of the local excitations in the atomic connectivity network determines the macroscopic viscosity of a liquid at high temperatures. The local configurational excitations (LCE's) in the atomic connectivity network are the elementary excitations in liquids, and their competition against phonons determines the cross-over phenomenon in liquids. The atomistic picture of liquid flow thus attained illustrates the importance of atomic dynamics in liquid flow, and will impact the way we perceive and describe the flow of liquid.

We thank S. Sastry, S. Yip, J. S. Langer, J. Bellissard, K. Kelton and J. R. Morris for useful discussions. The work was supported by the U.S. Department of Energy, Office of Basic Energy Sciences, Materials Science and Engineering Division. *Ab intio* calculations used resources of the National Energy Research Scientific Computing Center, which is supported by the Office of Science of the U.S. Department of Energy under Contract No. DE-AC02-05CH11231.

Table 1  Values of $\tau_\infty$ and activation energy at temperatures above $T_A$, $E_a$, of $\tau_M$ for three models. The KA model has a different energy scale.

| Model | $\tau_\infty$ (fs) | $E_a$ (meV) |
|---|---|---|
| Fe | 15.3 | 368 |
| Cu-Zr | 19.3 | 248 |
| Zr-Cu-Al | 23.0 | 222 |

**Figure captions:**

**Figure 1.** Bond lifetime, $\tau_B$, $\alpha$-relaxation time, $\tau_\alpha$, Maxwell relaxation time, $\tau_M$, and the lifetime of local atomic connectivity, $\tau_{LC}$, calculated for liquid iron as a function of $T_g/T$, where $T_g = 950K$ and $T$ is temperature. See text for the definition of each relaxation time.

**Figure 2.** Change in the local atomic connectivity by losing or gaining one nearest neighbor.

**Figure 3.** Temperature dependence of the Maxwell relaxation time, $\tau_M$, and the lifetime of local atomic connectivity, $\tau_{LC}$, calculated for various models of liquid as a function of $1/T$. $T_A$ is the crossover temperature above which the Arrhenius behavior is observed.

**Figure 4.** The ratio, $\tau_M/\tau_{LC}$, plotted against $T/T_A$ for various models, including the results of the *ab-initio* MD simulation for Cu-Zr. Above $T_A$, $\tau_M$ is approximately equal to $\tau_{LC}$ for all systems. $T_1$ is the temperature at which $\xi_p = c_T \tau_M$ becomes equal to $a$, the nearest neighbor distance. Agreement between $T_1$ and $T_A$ suggests that the crossover phenomenon is determined by the ability of crosstalk between the neighboring atoms by the phonon exchange.

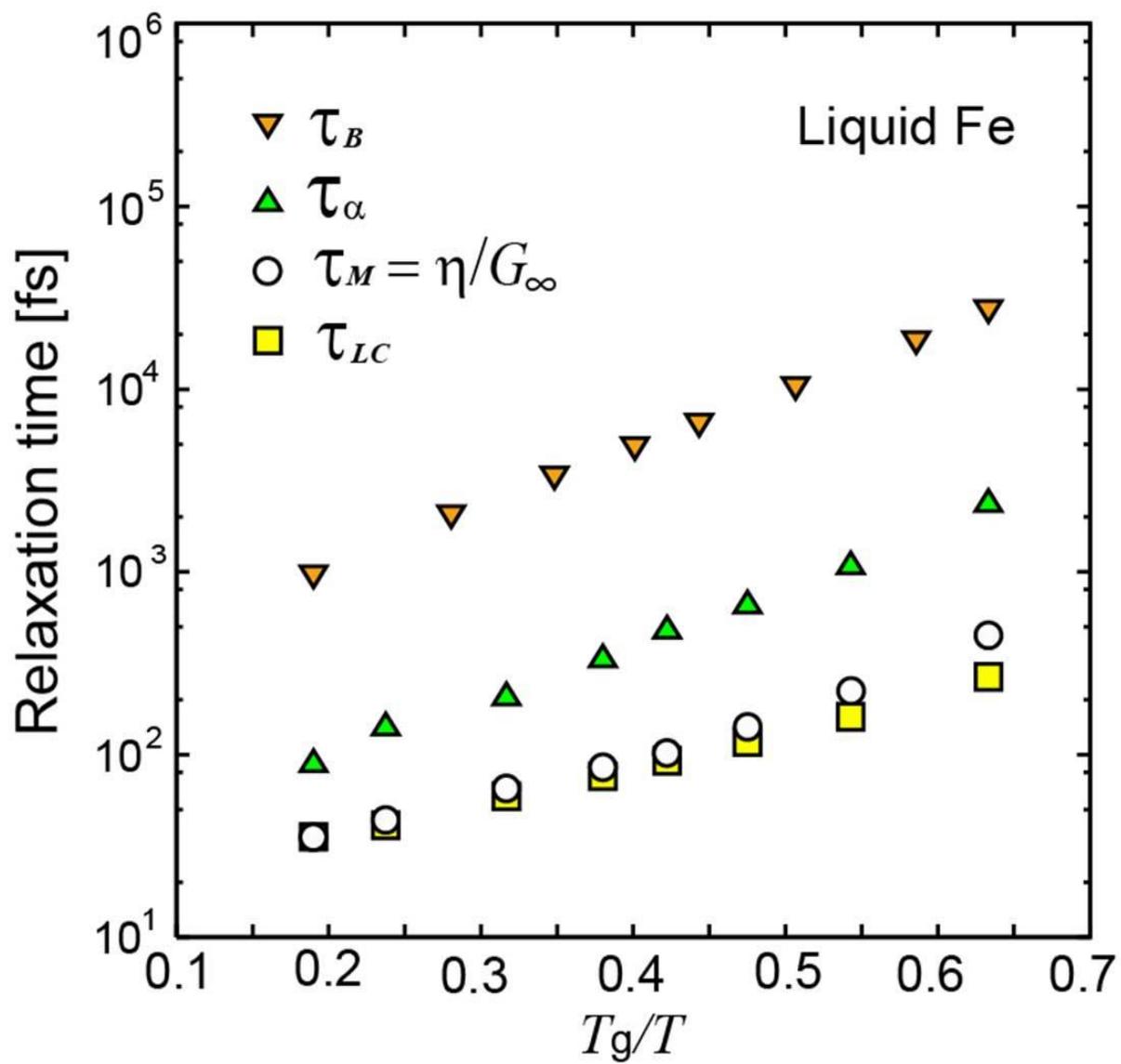

Fig. 1

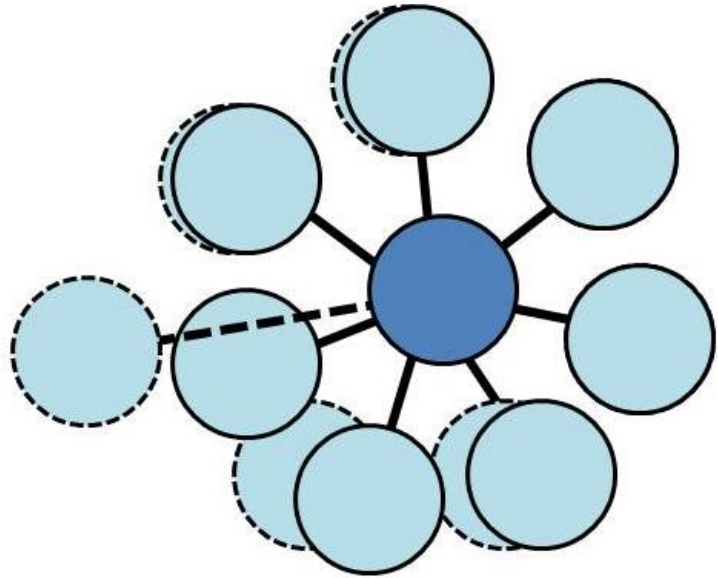

Fig. 2

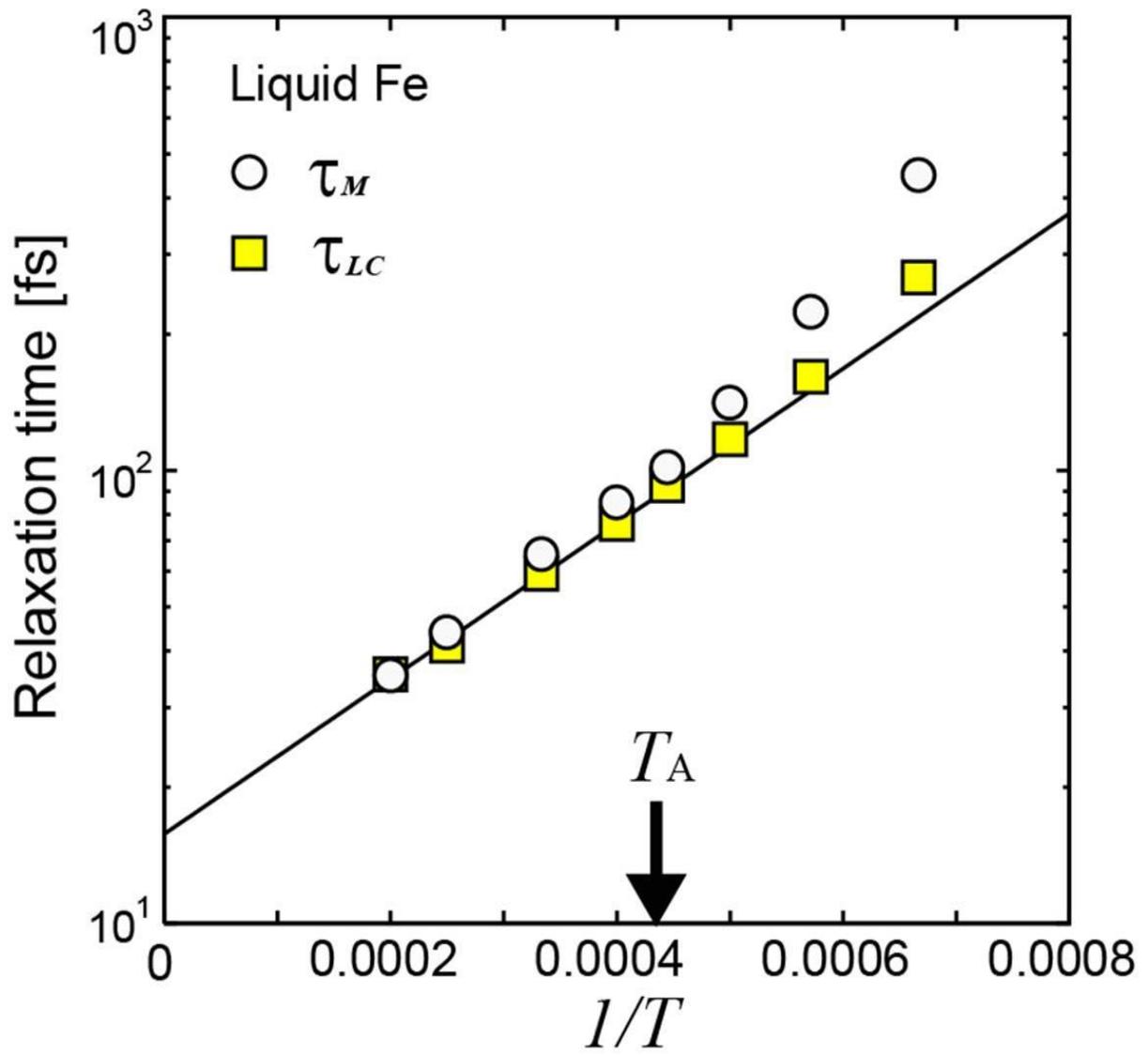

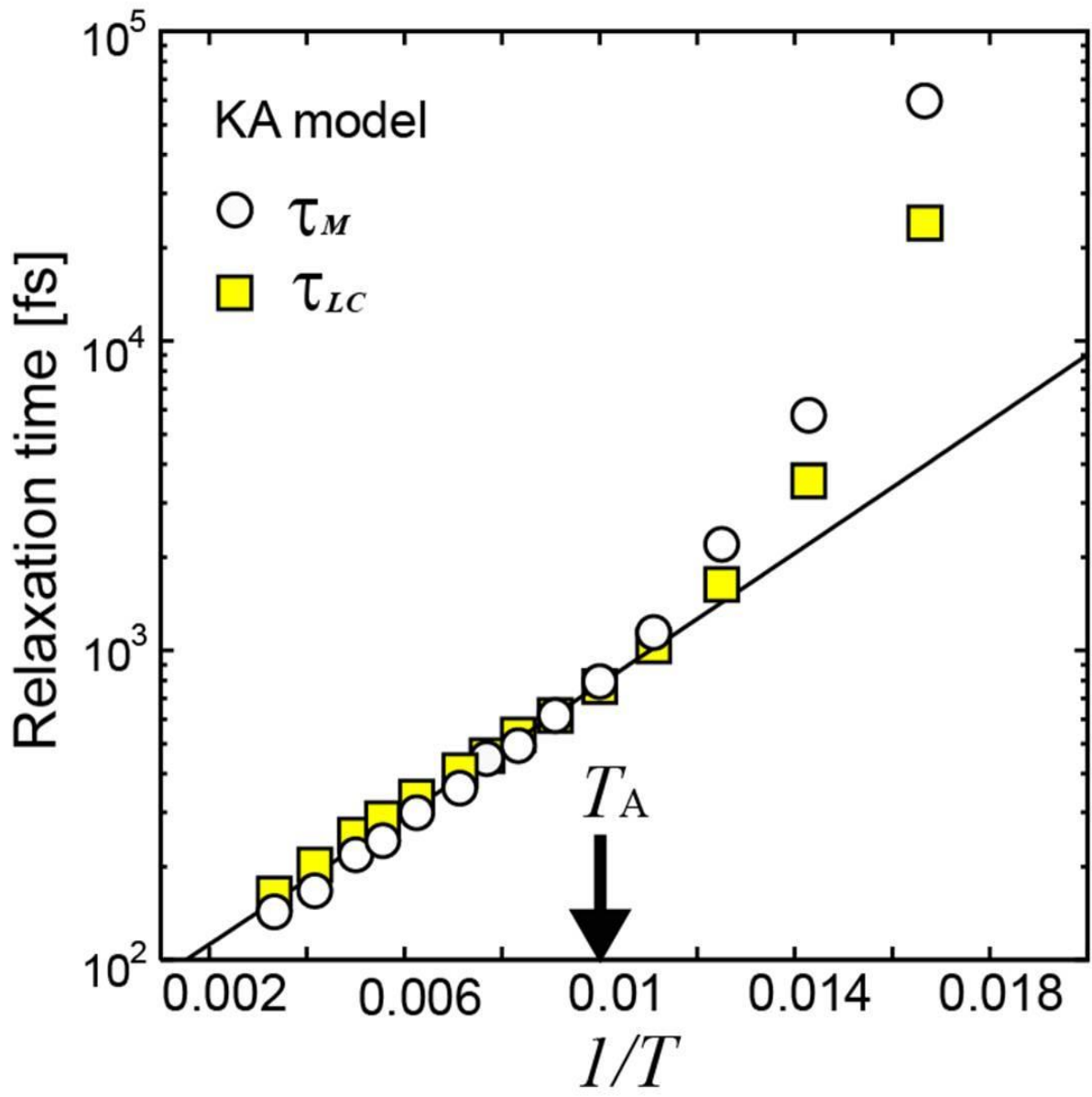

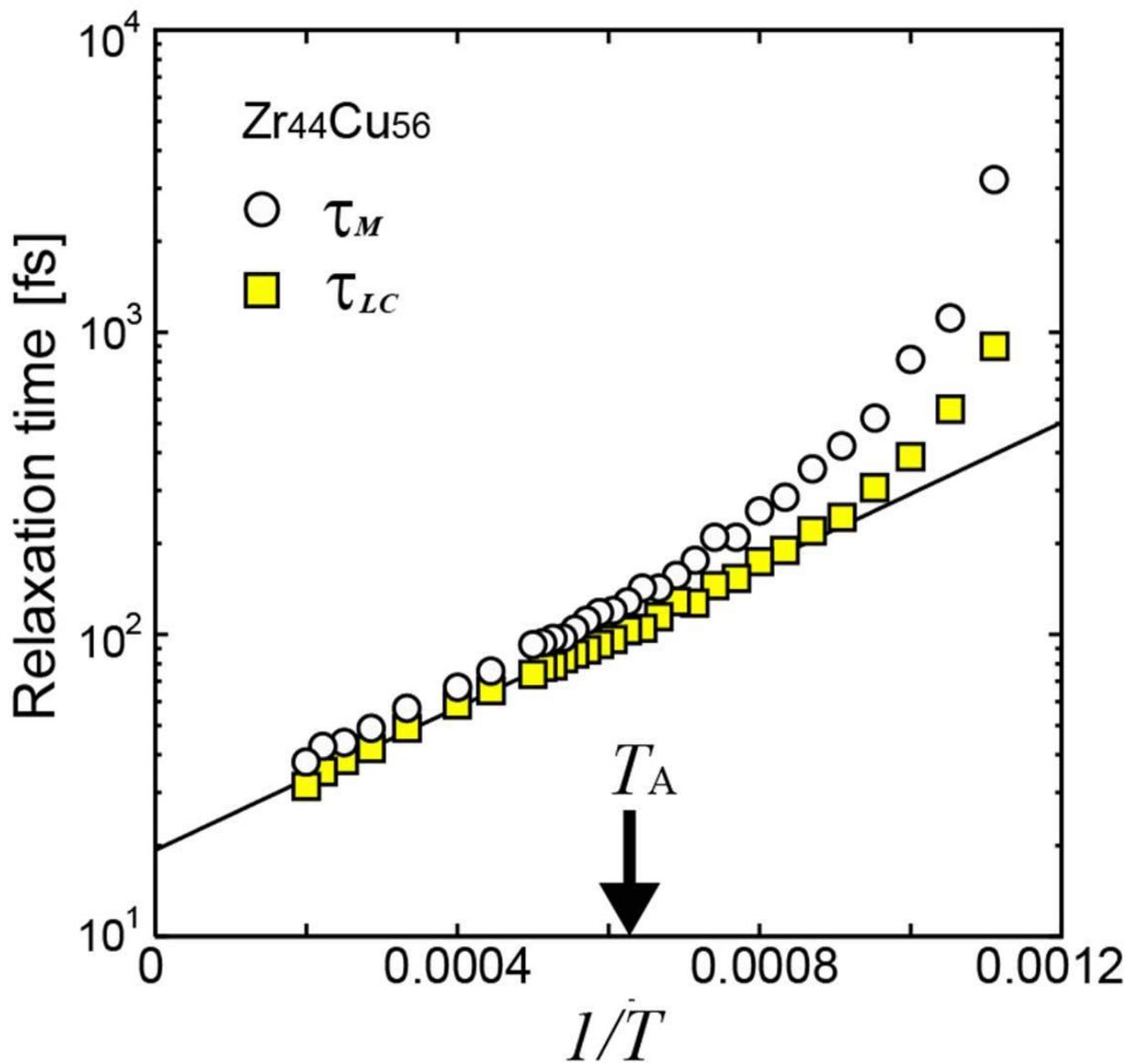

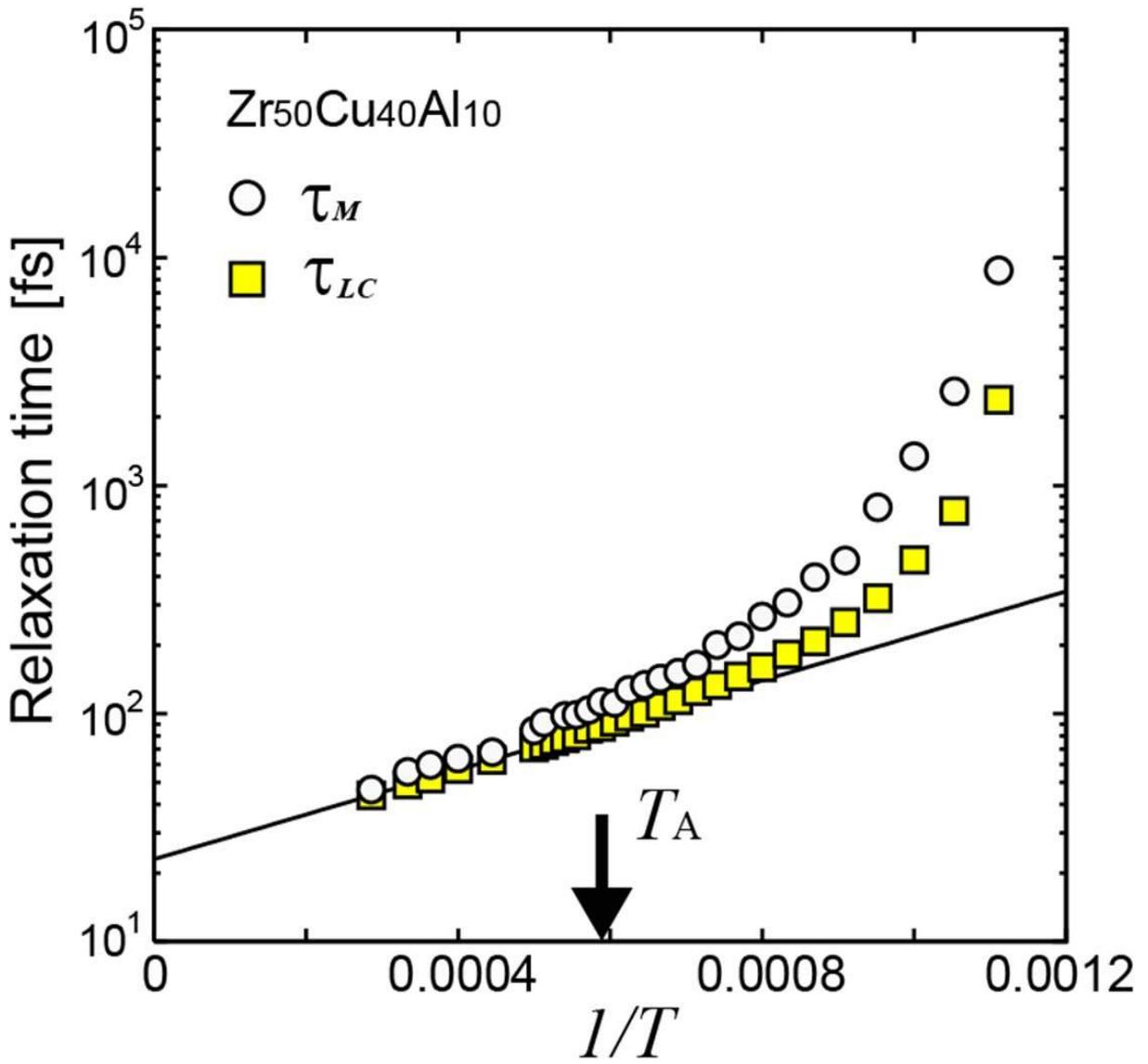

Fig. 3

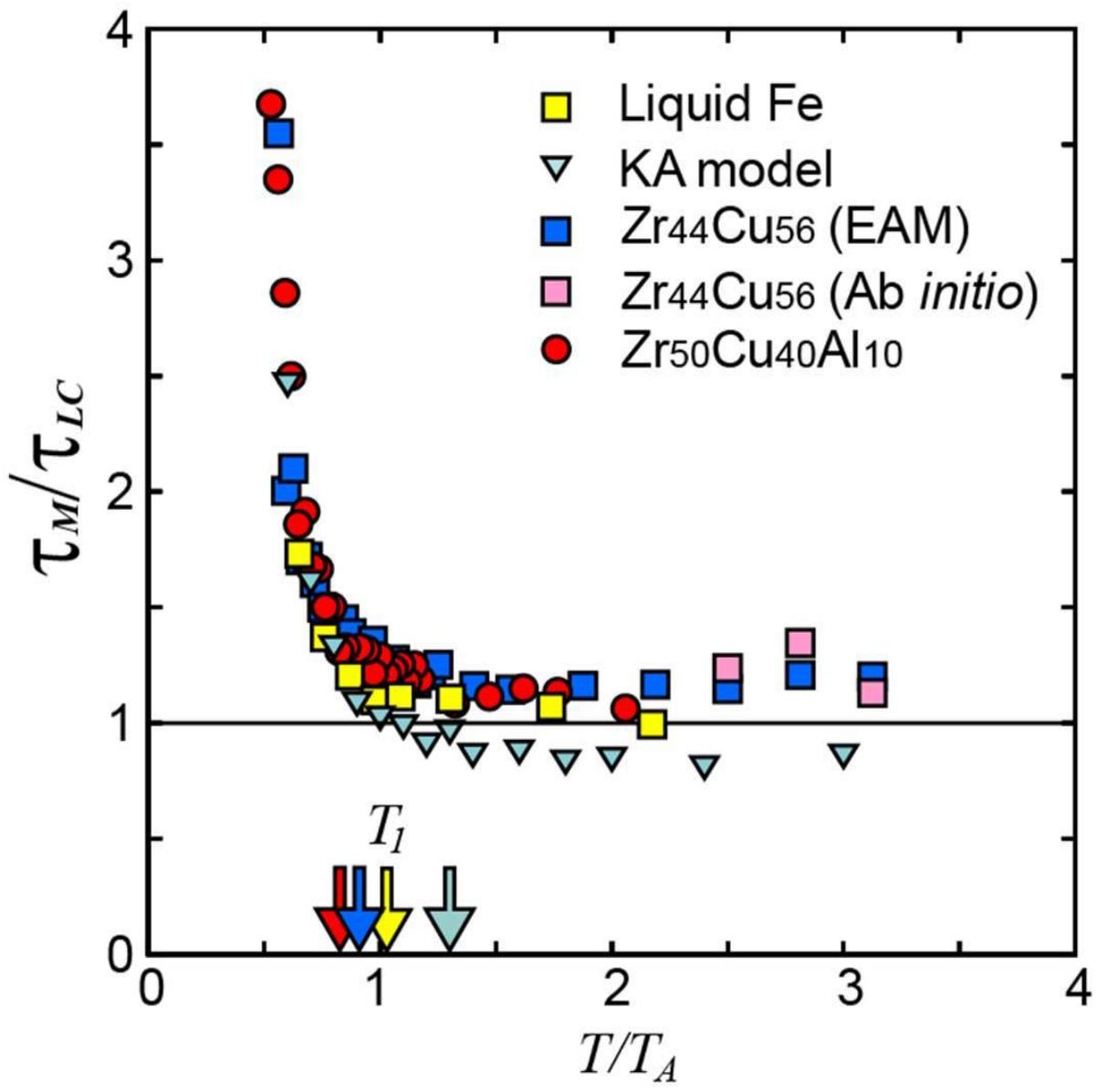

Fig. 4

# Supplementary Material

## 1. Details of classical and quantum mechanical molecular dynamics simulation

### A. Liquid iron

A monoatomic model of liquid Fe with the modified Johnson potential [9] consisted of 16,384 atoms. Simulations were performed in a temperature range from 1500K to 5000K, under NVT ensemble at a number density of 0.07843 Å$^{-3}$. For this system $T_g \sim$950K [S1] and $T_A \sim$2300K [29].

### B. Binary Kob-Andersen model

The Kob-Andersen binary alloy model [10] consisting of 10,976 Lennard-Jones (LJ) particles and the most popular composition, 80/20 mixture (80% particle A and 20% particle B), was chosen for simulation. All results are expressed in argon units. Both types of particles have the same mass, and the LJ interaction between two particles of type $\alpha$ and $\beta$ is given by $U_{\alpha\beta}(r) = 4\epsilon_{\alpha\beta}(\left(\frac{\sigma_{\alpha\beta}}{r}\right)^{12} - \left(\frac{\sigma_{\alpha\beta}}{r}\right)^{6})$; $\sigma_{AA}$=3.4Å, $\epsilon_{AA} = 120K\,k_B$, $\sigma_{AB} = 0.88\sigma_{AA}$, $\sigma_{BB} = 0.8\sigma_{AA}$, $\epsilon_{AB} = 1.5\epsilon_{AA}$, and $\epsilon_{BB} = 0.5\epsilon_{AA}$. The simulations were performed in a temperature range from 60K to 300K, under NVT ensemble at a reduced density of 1.2. $T_g$ of this system is 52.8 K [S2] and $T_A$ is reported to be around 100-120K [28, S3].

### C. $Cu_{56}Zr_{44}$

For $Cu_{56}Zr_{44}$ we employed the embedded atom method (EAM) potential [11] and simulated the systems with 16,000 atoms under NVT ensemble at a density of 7.3756 g/cm$^3$, or a number density of 0.05864 Å$^{-3}$. The simulations were carried out for high temperatures above $T_g$ using the LAMMPS software. From the jump in specific heat the glass transition temperatures were determined to be about 700K, which is in agreement with the value previously determined by simulations [S4].

### D. $Zr_{50}Cu_{40}Al_{10}$

For $Zr_{50}Cu_{40}Al_{10}$, we employed the embedded atom method (EAM) potential [12] and simulated the systems with 32,000 atoms under NTP ensemble, for which the pressure was maintained to be zero. At 2000K the number density was 0.05023 atoms/Å$^3$. The simulations were carried out for high temperatures above $T_g$ using the LAMMPS software. From the jump in specific heat the glass transition temperatures were determined to be about 860K for $Zr_{50}Cu_{40}Al_{10}$ [S5]. All simulation runs were long enough for the systems to reach the equilibrium state.

### E. Ab-initio MD for $Cu_{56}Zr_{44}$

*Ab initio* MD simulations were performed on a model of 200 atoms at a number density of 0.05864 Å$^{-3}$, using dynamics based on forces from the Projector Augmented-Wave (PAW) method [13] as implemented in the VASP 4.6 code [14]. At each time step the energy was converged to 10$^{-6}$ eV prior to force evaluation. The energy cut-off at 276.4 eV was checked by a further ps of data collection with an energy cut-off of 368.6 eV. A further 1.2 ps of data collection was performed with the temperature maintained by velocity scaling to confirm that the temperature fluctuations associated with the use of Nose-Hoover on small samples did not significantly affect the results. The Nose-Hoover-276.4 eV sample was used to initiate samples at T = 4000K, 4500K and 5000K by ramping the temperature over 0.25 ps followed by 1.0 ps and 1.25 ps of data collection respectively.

## 2. Atomic connectivity

The connectivity between a pair of atoms can be defined by atoms being the nearest neighbor to each other. The nearest neighbors are defined by the first minimum in the PDF, $r_{min}$; if a pair-distance between the atoms is less than $r_{min}$, then they are defined to be the nearest neighbors to each other, and atoms are topologically connected. It is possible to define such connectivity for dense simple liquids, because the PDFs usually show a clear separation between the 1$^{st}$ and 2$^{nd}$ peaks. Exceptions are for a pair of minority elements such as B-B pairs for the Kob-Andersen binary alloy and Al-Al pairs for $Zr_{50}Cu_{40}Al_{10}$, which do not show such a clear separation. For such pairs we chose a minimum between the pre-peak of the PDF, often observed in the partial PDF, and the maximum peak of the PDF. The effects of this minor uncertainty on main results are very small.

## 3. Lifetime of local atomic connectivity

The lifetime of local atomic connectivity is defined as follows. Firstly we define the atomic bonds for a configuration of atoms at $t = t_0$, and calculate the average coordination number, or the number of nearest neighbors. Usually the average coordination number is around 12 for dense simple liquids. Next by monitoring the atomic bonds present at $t = t_0$ we detect the breaking of some bonds for $t > t_0$, where their atomic distances become larger than $r_{min}$. This results in the decrease in the average coordination number with time. Then we can define the lifetime of local atomic connectivity and the bond-lifetime as $N_c(t_0 + \tau_{LC}) = N_c(t_0) - 1$ and $N_c(t_0 + t_B) = N_c(t_0)/e$, respectively.

## 4. Phonon localization

In the hydrodynamic theory the imaginary part of the shear modulus is given by $i\eta\omega = i\tau_M\omega G_\infty$. Thus the phonon linewidth is,

$$\Gamma = \text{Im}\left(\sqrt{\frac{G}{\rho}}\right) = \sqrt{\frac{G}{\rho}}\frac{\tau_M\omega}{2}. \quad (S1)$$

The Ioffe-Regel localization condition is $\Gamma/\omega > C_L$ [26]. Thus,

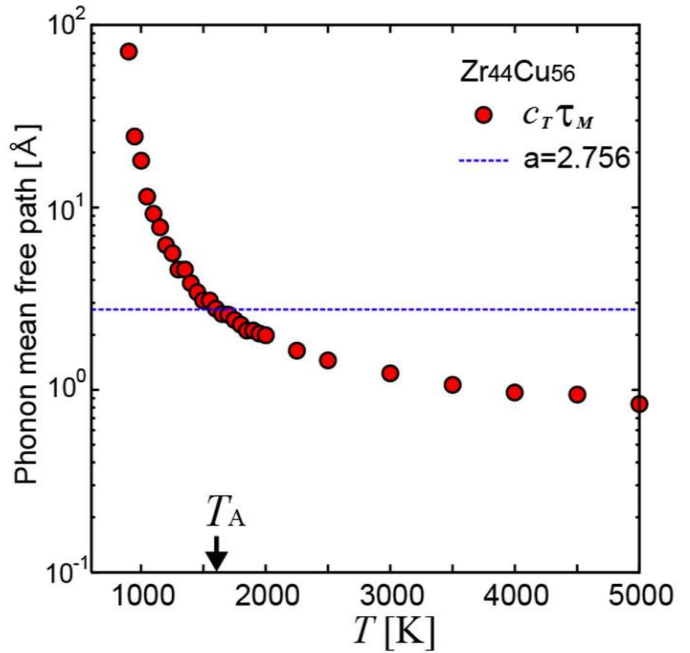

Fig. S1 Temperature dependence of the phonon mean-free path calculated for $Cu_{56}Zr_{44}$. The dashed line indicates the nearest neighbor distance. The crossing temperature, $T_I$, is close to $T_A$.

$$\frac{\Gamma}{\omega} = \frac{c_T \tau_M}{2} k = \frac{\pi c_T \tau_M}{\lambda} \tag{S2}$$

For the shortest phonon wave ($\lambda = 2a$) this gives $\Gamma/\omega = (\pi/2)(c_T \tau_M/a)$. Fig. S1 shows the plot of $\xi_p = c_T \tau_M$ against $T$ for $Cu_{56}Zr_{44}$. At $T_A$ $\xi_p \approx a$.